\documentclass{article}

\usepackage[final]{neurips_2023_ml4ps}

\usepackage[utf8]{inputenc} 
\usepackage[T1]{fontenc}    
\usepackage{hyperref}       
\usepackage{url}            
\usepackage{booktabs}       
\usepackage{amsfonts}       
\usepackage{nicefrac}       
\usepackage{microtype}      
\usepackage{xcolor}         
\usepackage{natbib}
\setcitestyle{square,sort,comma,numbers}
\usepackage{graphicx}

\title{MCMC to address model misspecification in Deep Learning classification of Radio Galaxies}

\author{%
  Devina Mohan \\
  Department of Physics \& Astronomy\\
  University of Manchester, UK\\
  \texttt{devina.mohan@postgrad.manchester.ac.uk} \\
  \And
  Anna Scaife\thanks{The Alan Turing Institute, 96 Euston Rd, London, UK \texttt{a.scaife@turing.ac.uk}} \\
  Department of Physics \& Astronomy\\
  University of Manchester, UK\\
  \texttt{anna.scaife@manchester.ac.uk} \\
}

\begin{document}

\maketitle

\begin{abstract}

The radio astronomy community is adopting deep learning techniques to deal with the huge data volumes expected from the next-generation of radio observatories. Bayesian neural networks (BNNs) provide a principled way to model uncertainty in the predictions made by deep learning models and will play an important role in extracting well-calibrated uncertainty estimates from the outputs of these models. However, most commonly used approximate Bayesian inference techniques such as variational inference and MCMC-based algorithms experience a "cold posterior effect (CPE)", according to which the posterior must be down-weighted in order to get good predictive performance. The CPE has been linked to several factors such as data augmentation or dataset curation leading to a misspecified likelihood and prior misspecification.  In this work we use MCMC sampling to show that a Gaussian parametric family is a poor variational approximation to the true posterior and gives rise to the CPE previously observed in morphological classification of radio galaxies using variational inference based BNNs.

\end{abstract}


\section{Introduction}

The next-generation of radio astronomy facilities such as the Square Kilometre Array (SKA) will produce huge volumes of data and the use of deep learning (DL) methods is inevitable given the expected data volumes \cite{scaife2020big, an2019science}. 
Modern astrophysics is driven by population analyses and any automated classification pipeline should produce well-calibrated uncertainty estimates that quantify the model uncertainty introduced in the results. In this work we consider the morphological classification of radio galaxies and discuss the challenges faced while implementing Bayesian Convolutional Neural Networks (CNNs) for their classification. 


While several works have looked at classifying radio galaxies with deep learning \cite[e.g.][]{Aniyan2017ClassifyingNetwork, lukic2019, tang2019, bowles2021attention, slijepcevic2022radio}, with the exception of \cite{scaife2021fanaroff} and \cite{mohan2022quantifying}, little work has been done on understanding the degree of confidence with which CNN models predict the class of individual radio galaxies. In general, it has been suggested that deep learning models produce overconfident predictions \citep{guo2017calibration} and provide no uncertainty estimates, which are essential for scientific application of these models. On the other hand, probabilistic models such as Bayesian neural networks (BNNs)
provide a principled way to model uncertainty \citep{mackay1992a,mackay1992b} by specifying priors, $P(\theta)$, over the neural network parameters, $\theta$, and learning the posterior distribution, $P(\theta|D)$, over those parameters, where $D$ is the data. 

Recovering this posterior distribution directly is intractable for neural networks. Several techniques have been developed to approximate Bayesian inference for neural networks among which Variational Inference (VI) and Monte Carlo (MC) Dropout methods are most commonly used. VI assumes an approximate posterior from a family of tractable distributions, and converts the inference problem into an optimisation problem \citep{practicalvi, blundell, vireview}.  The model learns the parameters of the distributions by minimising an Evidence Lower Bound Objective (ELBO) function.

Another easily implemented Bayesian approximation is MC Dropout, which learns a distribution over the network outputs by setting randomly selected weights of the network to zero with probability, $p$ \citep{gal2015bayesian}. MC dropout can be considered an approximation to VI, where the variational approximation is a Bernoulli distribution. Although a convenient technique, this method lacks flexibility and does not fully capture the uncertainty in model predictions, especially under covariate shift where the data distributions at training and test time are not identically distributed \citep{chan2020unlabelled}.

However, there are several challenges in implementing BNNs in practice. 
Several published works have reported that their BNNs experience a "cold posterior effect (CPE)", according to which the posterior needs to be down-weighted or tempered with a temperature term, $T \leq 1$, in order to get good predictive performance \citep{wenzel2020good}:
\begin{equation}
    P(\theta|D) \propto (P(D|\theta) P(\theta))^{1/T}.
    \label{eq:cpe}
\end{equation}





Previous work has shown that VI based BNN models experience a CPE when classifying radio galaxies \citep{mohan2022quantifying}.
The choice of variational approximation limits the variational posterior to specific regions of the true posterior density space and it is difficult to evaluate how good the variational approximation is without having access to the true posterior \citep{yao2018yes}.
Several hypothesis have been put forward to explain the CPE including likelihood, prior and model misspecification \citep{ wenzel2020good, izmailov2021bayesian, noci2021disentangling}. 

In this work we demonstrate that, for radio galaxy classification, using MCMC to recover posterior distributions on neural network parameters suggests that the "cold posterior effect" previously observed with VI models is due to model misspecification arising from poor variational posterior approximations. We also compare model performance for different approximate Bayesian inference methods including VI and MC Dropout and present preliminary uncertainty calibration results.




\section{MCMC for Neural Networks}

MCMC methods are a class of algorithms used to obtain samples from probability distributions which are otherwise intractable or do not have a full analytical description. 
The first application of MCMC to neural networks was proposed by \citep{neal1998view}, who introduced Hamiltonian Monte Carlo (HMC) from quantum chromodynamics to the general statistics literature. 
However, it wasn't until \citep{welling2011bayesian} introduced Stochastic Gradient Langevin Dynamics (SGLD), that MCMC for neural networks became feasible for large datasets. 
More recently, \citep{cobb2021scaling} have revisited HMC and proposed novel data splitting techniques to make it work with large datasets. We use the HMC algorithm in our work. 

\textbf{Hamiltonian Monte Carlo} HMC simulates the path of a particle traversing the negative posterior density space using Hamiltonian dynamics \citep{neal2011mcmc, betancourt2017conceptual,hogg2018data}. 
To apply HMC to deep learning, the neural network parameter space is augmented by specifying an additional momentum variable, $m$, for each parameter, $\theta$. Therefore, for a $d$-dimensional parameter space, the augmented parameter space contains $2d$ dimensions. We can then  define a log joint density as follows:
\begin{equation}
    \mathrm{log}[p(\theta, m)] = \mathrm{log}[p(\theta|D) p(m)] \, .
    \label{eq:joint_density}
\end{equation}
%

Hamiltonian dynamics allows us to travel on the contours defined by the joint density of the position and momentum variables, also known as the phase space. The Hamiltonian function is given by: 
\begin{equation}
    H(\theta, m) = U(\theta) + K(m) = constant ,
    \label{eq:ham}
\end{equation}
where $U(\theta)$ is the potential energy and $K(m)$ is the kinetic energy. The potential energy is defined to be the negative log posterior probability and the kinetic energy is usually assumed to be quadratic in nature and of the form $ K(m) = (1/2) \, m^{T} M^{-1} m$,
%
%
where $M$ is a positive-definite mass matrix. This corresponds to the negative probability density of a zero-mean Gaussian, $p(m) = \mathcal{N}(m|0, M)$, with covariance matrix, M, which is usually assumed to be the identity matrix. 

%

The partial derivatives of the Hamiltonian describe how the system evolves with time.
In order to solve the partial differential equations using computers, we need to discretise the time, $t$, of the dynamical simulation using a step-size, $\epsilon$. The state of the system can then be computed iteratively at times $\epsilon$, $2\epsilon$, $3\epsilon$... and so on, starting at time zero upto a specified number of steps, $L$.
The leapfrog integrator is used to solve the system of partial differential equations.
Two hyperparameters, the step-size, $\epsilon$, and the number of leapfrog steps, $L$, together determine the trajectory length of the simulation. 
The partial derivative of the potential energy with respect to the position, $\partial U/\partial \theta$, can be calculated using the automatic differentiation capabilities of most standard neural network libraries. 

In each iteration of the HMC algorithm, new momentum values are sampled from Gaussian distributions, followed by simulating the trajectory of the particles according to Hamiltonian dynamics for $L$ steps using the leapfrog integrator with step-size $\epsilon$. At the end of the trajectory, the final position and momentum variables, $(\theta^{*}, m^{*})$,  are accepted based on a Metropolis-Hastings accept/reject criterion that evaluates the Hamiltonian for the proposed parameters and the previous parameters. 

\section{Experimental Setup}

\textbf{Data}
The MiraBest dataset used in this work consists of 1256 images of radio galaxies of $150\times 150$ pixels pre-processed to be used specifically for deep learning tasks \citep{porter2023mirabest}. The galaxies are labelled using the FRI and FRII morphological types based on the definition of \citep{fanaroff1974morphology} and further divided into their subtypes. In addition to labelling the sources as FRI, FRII and their subtypes, each source is also flagged as `Confident' or `Uncertain' to indicate the human classifiers' confidence while labelling the dataset. In this work we use the MiraBest Confident subset and consider only the binary FRI/FRII classification. The training and validation sets are created by splitting the predefined training data into a ratio of 80:20. The final split consists of 584 training samples, 145 validation samples, and 104 withheld test samples. No data augmentation is used.


\textbf{Architecture}
We use an expanded LeNet-5 architecture with two additional convolutional layers with 26 and 32 channels, respectively, to be consistent with the literature on using BNNs for classifying the MiraBest dataset \citep{mohan2022quantifying}. The model has $232, 444$ parameters in total.

\textbf{MCMC Inference}
We use the \textsc{hamiltorch} package\footnote{\url{ https://github.com/AdamCobb/hamiltorch}} developed by \citep{cobb2021scaling} for scaling HMC to large datasets. 
Using their HMC sampler, we set up two HMC chains of $200,000$ steps using different random seeds and run it on the MiraBest Confident dataset. We use a step size of $ \epsilon = 10^{-4}$ and set the number of leapfrog steps to $L = 50$. We specify a Gaussian prior over the network parameters and evaluate different prior widths, $\sigma = \{ {1, 10^{-1},10^{-2}, 10^{-3} \}}$, using the validation data set. We find that $\sigma = {10^{-1} }$ results in the best predictive performance and consequently use it to define the prior width for all weights and biases of the neural network in our experiments. A burn-in of $20,000$ samples is discarded. To compute the final posteriors we thin the chains by a factor of 1000 to reduce the autocorrelation in the samples and obtain 180 samples. A compute time of $150$ hrs is required to run the inference on two Nvidia A100 GPUs. 
The Gelman-Rubin diagnostic, $\hat{R}$, is used to assess the convergence of our HMC chains \citep{gelman1992inference}. 
If $\hat{R} \approx 1$ we consider the MCMC chains for that particular parameter to have converged. While $\hat{R}$ values for some parameters in the network are greater than $1$, the final two neurons in the last layer of our network have $\hat{R} \leq 1$. We also monitor the negative log-likelihood and accuracy, which converge by the $100,000^{\textit{\emph{th}}}$ inference step.

\textbf{Other models} For the VI implementation we use a Gaussian variational approximation to the posterior and consider different priors including Gaussian and Laplace distributions following \citep{mohan2022quantifying}. The Laplace prior provides optimal predictive performance and lowest uncertainty calibration error, however for direct comparison to our HMC baseline we also consider a Gaussian prior with $\sigma = 0.01$. Results are reported for a tempered VI posterior, with $T = 0.01$ in Table~\ref{tab:model_performance}. For the MC Dropout model, a dropout rate of 50\%  is implemented before the last layer of our neural network, which is standard for CNNs \citep{scaife2021fanaroff, gal2015bayesian}. The network is trained for 150 epochs using the Adam optimser with a learning rate of $10^{-3}$ and a weight decay of $10^{-4}$. We obtain 200 samples from VI and MC Dropout posterior predictive distributions by passing each sample in the test set through the test loop 200 times. We use the same optimiser hyperparameters as the MC Dropout training for our non-Bayesian CNN model. A compute time of $12$ mins is required to train the VI model on a single Nvidia A100 GPU. 

Code for this work is available at \url{ https://github.com/devinamhn/RadioGalaxies-BNNs}

\section{Results}

\begin{figure*}
    \centering
    \includegraphics[width=\textwidth]{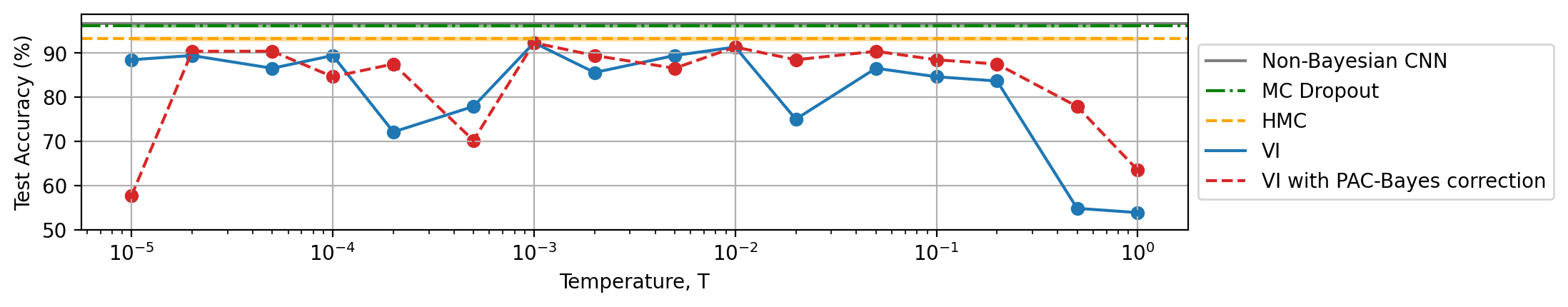}
    \caption{The “cold posterior” effect (CPE) observed in VI models (solid blue line) persists inspite of a PAC-Bayes correction term to account for model misspecification (red dashed line) when classifying radio galaxies. Using samples from HMC, we demonstrate that using a Gaussian variational distribution leads to a poor approximation to the posterior, giving rise to the CPE. Data are also shown for MC Dropout (green dot-dashed line) and our non-Bayesian CNN (solid gray line) for comparison.}
    \label{fig:CPE}
\end{figure*}

\textbf{Cold posterior effect} Previous work on using VI for radio galaxy classification has shown that the 
"cold posterior effect" (CPE) persists even when the learning strategy is modified to compensate for model misspecification with a second order PAC-Bayes bound to improve the generalisation performance of the network \citep{mohan2022quantifying, masegosa2019learning}, see Figure \ref{fig:CPE}. We do not observe a CPE when we use samples from our HMC inference to construct the posterior predictive distribution for classifying the MiraBest dataset (orange dashed line in Figure \ref{fig:CPE}). This suggests that using a Gaussian parametric family as a variational approximation to the true posterior distribution is a poor assumption and leads to a misspecified model, which gives rise to the CPE.  
In the general Bayesian DL literature, some authors argue that CPE is mainly an artifact of data augmentation \citep{izmailov2021bayesian}, while others have shown that data augmentation is a sufficient but not necessary condition for CPE to be present \citep{noci2021disentangling}. We find that data augmentation does not have a significant effect on our HMC and VI models.

\textbf{Model performance} The test error is calculated by taking an average of the predictions obtained using the expected value of the posterior predictive distribution for each galaxy in the MiraBest Confident test set for different models, see Table \ref{tab:model_performance}. No data augmentation is used during training/inference. The non-Bayesian CNN and MC Dropout perform comparably in terms of test error. HMC is more accurate than VI, but does not match the predictive performance of MC Dropout. We note that it is not performance alone that is important for our application, but also the calibration of the posterior uncertainties which will influence the scientific analysis performed using the catalogues generated by DL pipelines. We have conducted preliminary uncertainty calibration experiments using the $64\%$ credible intervals of the posterior predictive distributions to calculate the class-wise expected Uncertainty Calibration Error (cUCE) values for the predictive entropy
\citep{mohan2022quantifying, gal2015bayesian, laves2019well}. However, at this stage we do not draw any strong conclusions from the uncertainty quantification experiments, which will be considered more fully in our future work.

\begin{table}
\centering
\caption[Model performance]{Test error and class-wise expected uncertainty calibration error (cUCE) for predictive entropy are reported for different BNNs for the MiraBest Confident dataset.}
	\begin{tabular}{lcc}
    \textbf{Model} & \textbf{Error (\%)}  & \textbf{\% cUCE}\\ 
    \hline
    Non-Bayesian CNN & $3.33$ & $\times$ \\
    MC Dropout & $3.85 \pm 0.19$ & $9.75$ \\ 
    VI (Laplace prior) & $10.58 \pm 0.30$  & $14.26$   \\  
    VI (Gaussian prior) & $12.96 \pm 0.33$  & $30.05$  \\
    HMC & $6.73 \pm 0.25$ &  $13.17$ \
   	\end{tabular}
   \label{tab:model_performance}
\end{table}





\section{Conclusions}
Using samples from HMC, we find that the cold posterior effect previously observed in the morphological classification of radio galaxies using variational inference arises from using a misspecified parametric family to approximate the posterior. While MCMC does not provide the most computationally efficient framework for approximate Bayesian inference for neural networks, it produces asymptotically exact samples from the posterior which are useful for developing more accurate approximate Bayesian inference techniques for the radio galaxy classification problem in future.


\begin{ack}
AMS gratefully acknowledges support from an Alan Turing Institute AI Fellowship EP/V030302/1.
\end{ack}

\bibliography{references}

\clearpage


\end{document}